# A note on monitoring the ratio of two Weibull percentiles

Pasquale Erto [1]

**Abstract.** This note introduces a new Bayesian control chart to compare two processes by monitoring the ratio of their percentiles under Weibull assumption. Both in-control and out-of-control parameters are supposed unknown. The chart analyses the sampling data directly, instead of transforming them in order to comply with the usual normality assumption, as most charts do. The chart uses the whole accumulated knowledge, resulting from the current and all the past samples, to monitor the current value of the ratio. Two real applications in the wood industry and in the concrete industry give a first picture of the features of the chart.

## 1. Introduction

Statistical process monitoring techniques have moved far beyond the traditional field of engineering into management, environmental science, molecular biology, genetics, epidemiology, clinical medicine, finance, law enforcement, athletics (e.g., see Stoumbos et al. 2000). They have found applications different from traditional quality control of products, such as health care monitoring, detection of genetic mutation, credit card fraud detection and insider trading in stock markets (e.g., see Chatterjee and Qiu, 2009).

In both traditional and new contexts, the need to compare two processes arises, even independently of their being in statistical control or not. For example, products from two different reactors or sister plants or production lines must be compared to know if the two processes are making the same product simultaneously. The comparison is usually made in terms of the ratio of specific properties of the two processes.

The control of the ratio of two properties of interest arises as a statistical problem in many fields such as biology, physics and economy. Examples include ratios of dollars overpaid (underpaid) to dollars paid in Unemployment Insurance quality control (Spisak, 1990), ratios of Mendelian inheritance in genetics, ratios of mass to energy in nuclear physics, ratios of target to control precipitation in meteorology, and ratios of inventory in economics (Nadarajah, 2006).

A specific widely used example of control of ratios is the "stress-strength" model in the context of reliability. It describes the life of a component that has a random strength $Y$ and is subjected to random stress $X$. Here, it is required that always $X/Y < 1$, otherwise the component fails at the instant that the stress $X$ exceeds the strength $Y$.

Further examples are found in the wood industry, where it is a common practice to compare properties for lumber of the same or different dimension, grade and species. Because United States lumber standards are given in terms of population fifth percentiles, the ratio is often expressed in terms of the fifth percentiles of two strength distributions (Huang and Johnson, 2006). Since it is generally accepted that lumber main properties follow the Weibull distribution, confidence intervals for the ratio of Weibull percentiles are studied (Johnson et al., 2003; Verrill et al., 2012).

---

[1] Department of Industrial Engineering, University of Naples Federico II, Italy, ertopa@unina.it

In literature, there are many contributions regarding the distribution of the ratio of two random variables coming from the same family such as the Gaussian (Marsaglia, 1965; Hinkley, 1969; Korhonen and Narula, 1989), the Student's t (Press, 1969; Kappenman, 1971), the Weibull (Basu and Lochner, 1971), the Non-Central Chi-Squared (Hawkins and Han, 1986), the Gamma (Lee et al., 1979; Provost, 1989), the Logistic (Nadarajaha and Gupta, 2006), the Frèchet (Nadarajaha and Kotz, 2006), the Inverted Gamma (Ali et al., 2007) the Laplace (Nadarajah, 2007) and the elliptically symmetric Kotz-type (Nadarajah, 2012). However, to address the comparison of two processes, often not all the parameters of the involved random variables are equally important. Rather, in real-life applications, the variables are more effectively compared with respect to some of their "operative" characteristics, such as the percentiles, as in the above-cited wood industry or other industrial contexts (e.g., see Padgett and Spurrier, 1990).

In general, when two whichever processes must be compared continuously, the solution can be a monitoring chart of the ratio of specific key characteristics associated with the processes, whose variability are often modelled via skewed distribution as the Weibull one (Meeker and Hamada, 1995).

The Weibull distribution has been used in many fields such as medicine, engineering, biology, public health, epidemiology, economics, demography, criminology, sociology, and ecology. In the industry, it is widely used in the manufacturing process of materials. For example, the breaking strength of the products is monitored by randomly selecting samples from the production process and sending to a destructive test to measure the breaking stress. Usually, not only the underlined distribution is the Weibull but also the collected data sets have extremely small size (say, 2 or 3), because of the time and cost needed for testing the samples.

Very few papers address the control chart of ratios (e.g., see Hosono et al. 1982; Spisak, 1990; Ramirez-Beltran, 1996). On the other hand, classical Shewhart control charts stand under normality assumption, so they are not effective when the distributions are skewed and the small size of the available samples prevents the use of the "normalizing effect" (see Shore, 2004). The case of the (skewed) Log-Normal distribution is an exception since the logarithm transformation of the observed data normalizes them. Moreover, monitoring the process mean and variance by means of classical control charts is less effective than monitoring percentiles since a slight variation in the mean and/or variance can produce a significant shift in the percentiles (see Padgett and Spurrier, 1990). Often, a specific percentile is considered a minimum threshold for reliability design and is more meaningful than other parameters such as the scale parameter or the mean (see, e.g., Bao et al., 2007). Therefore, the objective of this paper is to develop a chart to monitor the ratio of percentiles from two distinct processes under Weibull assumption. The paper starts from the results in (Erto and Pallotta 2007) and (Erto et. al., 2014) where Shewhart-type Bayesian control chart is used to monitor the Weibull percentiles of a *single* process. The former chart (Erto and Pallotta, 2007) combines the given initial prior information exclusively with the *current* sample data, and uses *fixed* control limits; it shows a detection power higher among others for Weibull processes as demonstrated in (Hsu et al., 2011). The latter chart (Erto et. al., 2014) combines prior information with *all* accumulated past and present sample data; it uses *updating* control limits from period to period and shows a detection power even higher than the former one.

The main features of the here proposed chart are: the ratio points on the chart are *cumulative* estimates, instead of *single-sample* estimates; the control limits are *continually updated* and can be calculated since the very first sample; it can be applied as an *acceptance control chart* based on eventually given specification limits.

The remainder of the paper is organized as follows. The next section introduces a specific re-parameterized form of the Weibull distribution suitable for the Bayesian approach. The following section describes the steps to draw the chart. Two real applicative examples, regarding the wood industry and the concrete industry, are reported in the last section. Some results given in (Erto and Pallotta, 2007) and (Erto et al., 2014), that are essential for the comprehension of this note, are briefly reported too.

## 2. The control charts for Weibull percentiles and their ratio

The Weibull distribution possesses many useful properties: it has only two parameters; its form is flexible to model different shapes; it has a simple likelihood function. Given a Weibull random variable $x$, the corresponding cumulative density function is:

$$F(x;\delta,\beta) = 1 - \exp\left[-(x/\delta)^{\beta}\right] \qquad x \geq 0; \quad \delta, \beta > 0 \tag{1}$$

where $\delta$ is the scale parameter and $\beta$ is the shape parameter.

Let $R$ denote a specified reliability level, usually chosen in the $0.90 - 0.99$ range, as reference level used in most industries (Meeker and Hamada 1995). Then, the corresponding Weibull percentile $x_R$ can be expressed as $x_R = \delta \left[\ln(1/R)\right]^{1/\beta}$ by setting (1) equal to $1-R$ and solving for $x_R$.

Often, in engineering, on the one hand, a specific Weibull percentile, $x_R$, is a critical quality characteristic of the products and on the other, there is prior information (e.g.: design data, specifications, previous experiments, expert opinion etc.) which can be reasonably quantified in terms of: 1) a range $(\beta_1, \beta_2)$ of the shape parameter (wide enough in order to plausibly contain the unknown true value); 2) an anticipated and reasonable value $\bar{x}_R$ for a percentile of the sampling distribution (Erto, 1982).

For these reasons, the re-parameterized form of the cumulative density function (1) in terms of the percentile $x_R$ and the shape parameter $\beta$:

$$F(x; x_R, \beta) = 1 - \exp\left[-\ln(1/R)(x/x_R)^{\beta}\right]; \quad x \geq 0; \quad x_R, \beta > 0; \tag{2}$$

is adopted, since it is a more suitable basis for the control chart developed in this section. For the $\beta$ parameter the Uniform prior pdf is assumed in the interval $(\beta_1, \beta_2)$, and the prior probability density function of $x_R$ is assumed to be the Inverse Weibull (Erto, 1982):

$$\text{pdf}\{x_R\} = ab(ax_R)^{-(b+1)} \exp\left[(ax_R)^{-b}\right]; \qquad x_R \geq 0; \qquad a,b > 0 \tag{3}$$

where $a$ and $b$ are the scale and shape parameters respectively, that must be defined. For $b$ the simplest choice is assuming $b = \beta$, as suggested in Erto (1982). In fact, the greater the shape parameter $\beta$ is, the more peaked the Weibull pdf is, the smaller the uncertainty in $x_R$ is, and then greater $b$ must be. This assumption leaves $b$ unspecified, and the practice shows that it works and it is preferable rather than to choose a fixed value, say $b = 3$, which often includes other information that is not actual prior knowledge about the prior. For $a$, since the expected value of (3) is:

$$\mathrm{E}\{x_R\} = \frac{\Gamma(1-1/b)}{a}. \qquad (4)$$

the prior value can be evaluated as:

$$a = \frac{\Gamma(1-1/\bar{b})}{\bar{x}_R} \qquad (5)$$

where $\bar{x}_R$ is the anticipated value for $\mathrm{E}\{x_R\}$, and $\bar{b}$ is the value of $b$ that we can anticipate on the basis of the prior interval $(\beta_1, \beta_2)$ for $\beta$ (coherently with the above assumption $b = \beta$):

$$\bar{b} = (\beta_1 + \beta_2)/2. \qquad (6)$$

The only restriction is $\beta_1 + \beta_2 > 2$ since $\beta_1 + \beta_2$ is used to set up the argument of the Gamma function in (4). The joint prior probability density function (pdf) of $x_R$ and $\beta$ is:

$$\mathrm{pdf}\{x_R, \beta\} = (\beta_2 - \beta_1)^{-1} a \beta (a x_R)^{-(\beta+1)} \exp\left[(a x_R)^{-\beta}\right] \qquad (7)$$

which combines the Uniform prior pdf for $\beta$ and the Inverse Weibull prior pdf for $x_R$ (3).

Consider the first process, that we want to monitor, and the first random vector $\underline{x}_1$ of $n$ data sampled from it. The corresponding likelihood function, under the Weibull assumption, is:

$$\mathrm{L}\left(\underline{x}_1 | x_R, \beta\right) \propto \beta^n x_R^{-\beta n} \prod_{i=1}^{n} x_i^{\beta-1} \exp\left[-x_R^{-\beta} \ln\left(\frac{1}{R}\right) \sum_{i=1}^{n} x_i^{\beta}\right] \qquad (8)$$

The Bayes' theorem combines the prior distribution (7) with the likelihood function (8) to obtain the first joint posterior pdf of $x_R$ and $\beta$ based on the first $n$ data. Recursively, at the $k^{\text{th}}$ sample $\underline{x}_k$, the joint posterior pdf results to be based on all the samples of size $n$ taken until then:

$$\mathrm{pdf}\{x_R, \beta | \underline{x}_1, \ldots, \underline{x}_k\} = \frac{\beta^{k \cdot n+1} a^{-\beta} x_R^{-\beta(k \cdot n+1)-1} \prod_{i=1}^{k \cdot n} x_i^{\beta-1} \exp\left[-x_R^{-\beta} A(k)\right]}{\Gamma(k \cdot n + 1) \int_{\beta_1}^{\beta_2} \beta^{k \cdot n} a^{-\beta} \prod_{i=1}^{k \cdot n} x_i^{\beta-1} A(k)^{-(k \cdot n+1)} d\beta} \qquad (9)$$

$$A(k) = a^{-\beta} + \ln\left(\frac{1}{R}\right) \sum_{i=1}^{k \cdot n} x_i^{\beta}$$

where the value of the prior parameter $a$ is obtained by replacing $\bar{x}_R$ in (5) with $\hat{x}_{R,k-1}$, the point estimate (12) of the $x_R$ percentile from the previous $(k-1)^{\text{th}}$ sample. Beside, using the following estimate (10) of the shape parameter $\beta$ from the previous $(k-1)^{\text{th}}$ sample, it is adopted $\beta_1 = \hat{\beta}_{k-1}/2$ and $\beta_2 = \hat{\beta}_{k-1} \times 1.5$ in order to obtain a reasonable large symmetrical interval.

The equation (9) represents both the posterior distribution after the $k^{\text{th}}$ sample and the prior for the $(k+1)^{\text{th}}$ one. These conjugate distributions stand for all sampling stages $(k \geq 1)$, but differ from the initial prior (7) $(k = 0)$. Note that the equation (9) is dif-ferent from that proposed in Erto and Pallotta (2007) and identically reported in Hsu et al. (2011).

After the $k^{\text{th}}$ sample, from equation (9) we can obtain the marginal posterior pdf for $\beta$, $\text{pdf}\{\beta|\underline{x}_k\}$, by integrating over $x_R$. From this we obtain the estimate of the shape parameter $\beta$ as the following *posterior* expectation that we can easily compute numerically:

$$\hat{\beta}_{x,k} = \text{E}\{\beta|\underline{x}_1,\ldots,\underline{x}_k\} = \frac{\int_{\beta_1}^{\beta_2} \beta^{k \cdot n+1} a^{-\beta} \prod_{i=1}^{k \cdot n} x_i^{\beta-1} A(k)^{-(k \cdot n+1)} d\beta}{\int_{\beta_1}^{\beta_2} \beta^{k \cdot n} a^{-\beta} \prod_{i=1}^{k \cdot n} x_i^{\beta-1} A(k)^{-(k \cdot n+1)} d\beta}. \quad (10)$$

This estimator can be used to check the stability of the Weibull shape parameter $\beta$. In fact, this parameter is a characteristic of the phenomenon under consideration. For example, it is related to the dispersion of flaws in the material (e.g., see Padgett et al. 1995). Thus $\beta$ can be considered constant, even if unknown, as in Nelson (1979).

Considering the second process that we want to monitor, and a random vector $\underline{y}_k$ of $n$ data sampled from it. Under the same assumptions made for the first process, we similarly obtain a second point estimate of the $\beta$, from the *posterior* expectation $\hat{\beta}_{y,k} = \text{E}\{\beta|\underline{y}_1,\ldots,\underline{y}_k\}$ like the (10).

Since $x_R$ is highly dependent on $\beta$, a conditional posterior pdf for $x_R$ is considered (Erto et al., 2014):

$$\text{pdf}\{x_R|\underline{x}_1,\ldots,\underline{x}_k,\beta\} = \frac{\beta\, x_R^{-\beta(k \cdot n+1)-1}}{\Gamma(k \cdot n+1)} A(k)^{k \cdot n+1} \exp\left[-x_R^{-\beta} A(k)\right] \quad (11)$$

The posterior expectation of (11) is adopted as point estimate of $x_R$:

$$\hat{x}_{R,k} = \text{E}\{x_R|\underline{x}_1,\ldots,\underline{x}_k,\overline{\beta}_k\} = \frac{\Gamma(k \cdot n+1-\overline{\beta}_k^{-1})}{\Gamma(k \cdot n+1)}\, A(k)^{\frac{1}{\overline{\beta}_k}} \quad (12)$$

where $\beta$ is replaced by:

$$\overline{\beta}_k = \frac{1}{k}\sum_{i=1}^{k}\left(\hat{\beta}_{x,i}+\hat{\beta}_{y,i}\right)\big/2 \quad (13)$$

which is the average of all the posterior estimates accumulated up to and including the two $k^{\text{th}}$ ones. This average estimate complies with the usual stability feature of $\beta$ and it is conservative, since also the calculation for each $\hat{\beta}_{x,i}$ and $\hat{\beta}_{y,i}$ uses the information from *all* previous samples. Note that the estimator (12) coincides with one of those proposed in (Erto et. al., 2014) but conceptually differs from that proposed in (Erto and Pallotta, 2007) and reported in Hsu et al. (2011) which, moreover, is not expressible in closed form.

Considering the second process, we similarly obtain the point estimate of the $y_R$ percentile:

$$\hat{y}_{R,k} = \text{E}\{y_R|\underline{y}_1,\ldots,\underline{y}_k,\overline{\beta}_k\} = \frac{\Gamma(k \cdot n+1-\overline{\beta}_k^{-1})}{\Gamma(k \cdot n+1)}\, B(k)^{\frac{1}{\overline{\beta}_k}}$$

$$B(k) = a^{-\overline{\beta}_k} + \ln\left(\frac{1}{R}\right)\sum_{i=1}^{k \cdot n} y_i^{\overline{\beta}_k} \quad (14)$$

where, as before, $\bar{\beta}_k$ is the average of all the posterior estimates of $\beta$ (13) accumulated up to and including the two $k^{\text{th}}$ ones.

From a technical point of view, the adopted approach can be considered a compromise between a pure Bayes and an empirical Bayes approach. In fact, to start the chart it is necessary to anticipate the values of the three hyper-parameters $\beta_1$, $\beta_2$ and $\bar{x}_R$ (in which prior engineers' knowledge can be really converted); the fourth hyper-parameter $a$ is deduced by (5) and the fifth hyper-parameter $b$ is left *unspecified*. After that, each sampled data set contributes to provide the values of all the hyper-parameters of the prior distributions for the analysis of the next sampled data set.

Once the estimates $\hat{x}_{R,k}$ (12) of the Weibull percentiles are obtained, we can plot them on the control chart of the first process. After the Phase I, they are compared with the control limits obtained as in the next step.

As anticipated, if a suspect about a shift in the Weibull shape parameter arises, also the control of the parameter $\beta$ can be performed by using the posterior estimates (10). This second control chart can be built together with the one for the percentile $x_R$.

The conditional random variable $x_R | \underline{x}_1, \ldots, \underline{x}_k, \bar{\beta}_k$ can be transformed into the $z$ standard Gamma random variable by using the transformation:

$$z = x_R^{\bar{\beta}_k} A(k); \qquad A(k) = a^{-\bar{\beta}_k} + \ln\left(\frac{1}{R}\right) \sum_{i=1}^{k \cdot n} x_i^{\bar{\beta}_k} \tag{15}$$

and the $\text{pdf}\{x_R | \underline{x}_1, \ldots, \underline{x}_k, \bar{\beta}_k\}$, obtained from (11) for $\beta = \bar{\beta}_k$:

$$\text{pdf}\{z\} = \frac{z^{k \cdot n}}{\Gamma(k \cdot n + 1)} \exp(-z). \tag{16}$$

Thus, using the inverse of the transformation (15):

$$x_R = z^{-1/\bar{\beta}_k} \left[ A(k) \right]^{1/\bar{\beta}_k} \tag{17}$$

and, given a false alarm risk $\alpha$, we can estimate the control limits $\text{LCL} = x_{R,\alpha/2}$ and $\text{UCL} = x_{R,1-\alpha/2}$ as simple transformations of the percentiles, $z_{1-\alpha/2}$ and $z_{\alpha/2}$ respectively, of the standard Gamma (16). These percentiles are re-calculated at every sample and their values obtained at the end of the Phase I become the control limits of the chart.

An analogous procedure must be followed to plot the control chart for the second process, starting from the estimates $\hat{y}_{R,k}$ (14) of the Weibull percentiles.

Assuming that the two percentiles, $x_R$ and $y_R$, are independent random variables with the same $\beta$ shape parameter (as usually found in practice), the probability density function of their ratio $u = x_R / y_R$ is:

$$\text{pdf}\{u | \underline{x}_1, \ldots, \underline{x}_k, \underline{y}_1, \ldots, \underline{y}_k, \beta\} = \beta \frac{\Gamma[2(k \cdot n + 1)]}{[\Gamma(k \cdot n + 1)]^2} u^{\beta(k \cdot n + 1) - 1} \frac{[C(k)]^{k \cdot n + 1}}{[1 + u^\beta C(k)]^{2(k \cdot n + 1)}}. \tag{18}$$

$$C(k) = B(k)/A(k)$$

Using the one-to-one transformation:

$$v = u^{\bar{\beta}_k} C(k) \tag{19}$$

where $\beta$ is replaced by $\bar{\beta}_k$ (13), we obtain the probability density function of $v$:

$$\text{pdf}\{v\} = \frac{\Gamma[2(k \cdot n + 1)]}{[\Gamma(k \cdot n + 1)]^2} \frac{v^{k \cdot n}}{[1 + v]^{2(k \cdot n + 1)}} \tag{20}$$

which is the Inverted Beta. Thus, using the inverse of the transformation (19) and given a false alarm risk $\alpha$, we can easily estimate the control limits $LCL = u_{R,\alpha/2}$ and $LCL = u_{R,1-\alpha/2}$ of the $u = x_R/y_R$ ratio as simple transformations of the percentiles, $v_{\alpha/2}$ and $v_{1-\alpha/2}$ respectively, of the Inverted Beta.

## 3. Example: control of modulus of rupture of fir specimens

This section shows an introductory application of the proposed control chart to the data given in Huang and Johnson (2006) to compare two productions of specimens of the same (Douglas) fir tree but with different cross sections, $2\times4$ and $2\times6$ inches respectively. The parameter of interest is the 0.05 percentile of the distribution of the modulus of rupture (MOR). It is generally expected that the MOR follows the Weibull distribution (Johnson et al., 2003; Verrill et al., 2012) with a high $\beta$ parameter, say $\beta = 5$, since a close to symmetrical shape is expected in this case (Huang and Johnson 2006).

From the first manufacturing process, a sample of $n = 4$ specimens, each with cross-section of $2\times4 = 8$ square inches, is selected periodically and the MOR of each specimen is measured in GPa (Giga-Pascals) and reported in Table 1 divided by 10. For these specimens the percentile value $\bar{x}_{0.95} = 2.9$ (GPa×10) is anticipated for $\bar{x}_R$ (5) and the prior interval $5.0\times(1\mp0.5)$ (i.e.: 2.5, 7.5) is anticipated for $\beta$.

Table 1. First process: $x$ MOR (GPa×10) of samples of $n = 4$ specimens with 2×4 inches cross section.

| | | | | |
|---|---|---|---|---|
| 3.7, 3.3, 4.9, 4.3 | 4.8, 4.6, 5.6, 4.7 | 4.8, 4.0, 4.6, 4.2 | 5.2, 4.4, 5.3, 5.0 | 4.8, 3.1, 3.9, 4.3 |
| 4.5, 4.2, 3.8, 4.1 | 3.2, 3.0, 3.6, 5.6 | 3.2, 2.4, 3.6, 4.0 | 3.0, 6.4, 4.2, 2.8 | 3.9, 3.8, 3.4, 2.5 |
| 2.9, 1.7, 3.4, 2.9 | 3.3, 3.7, 4.0, 3.3 | 4.1, 3.8, 4.4, 1.9 | 3.1, 3.7, 3.9, 2.7 | 3.1, 2.7, 3.5, 2.8 |
| 3.5, 3.9, 3.2, 4.1 | 1.6, 1.9, 3.8, 2.6 | 2.2, 3.4, 1.6, 1.8 | 3.3, 2.1, 2.9, 3.0 | 2.1, 3.6, 2.4, 3.1 |
| 2.3, 2.2, 3.6, 2.9 | 2.7, 1.9, 3.1, 3.4 | 3.5, 3.6, 0.98, 2.1 | 3.1, 1.3, 2.5, 2.3 | 4.7, 1.8, 0.85, 4.1 |

The first ten samples (of size $n = 4$) are supposed in-control, and are used in the Phase I. The percentile values $\hat{x}_{0.95}$ are recursively estimated by (12).

From the second process, a sample of $n = 4$ specimens, each with cross-section of $2\times6 = 12$ square inches, is selected from the manufacturing process periodically, and the MOR of each specimen is measured in GPa (Giga-Pascals) and reported in Table 2 divided by 10. For these specimens a different 0.05 percentile value, $\bar{y}_{0.95} = 3.8$ (GPa×10), is anticipated (depending on the different cross section) but the same prior interval for $\beta$ (i.e.: 2.5, 7.5) is anticipated. As before, the first ten samples (of size $n = 4$) are supposed in-control, and are used in the Phase I. The percentile values $\hat{y}_{0.95}$ are recursively estimated by (14).

Table 2. Second process: $y$ MOR (GPa×10) of samples of $n = 4$ specimens with 2×6 inches cross section.

| | | | | |
|---|---|---|---|---|
| 6.6, 4.5, 5.8, 6.5 | 6.4, 7.3, 5.6, 6.8 | 5.5, 5.7, 5.4, 5.5 | 6.2, 5.3, 4.6, 6.0 | 7.6, 6.3, 5.8, 7.1 |
| 6.1, 4.7, 5.4, 4.6 | 5.4, 3.5, 4.5, 4.5 | 3.8, 5.0, 5.5, 4.9 | 6.2, 6.2, 5.8, 5.3 | 4.7, 5.7, 4.6, 5.4 |
| 5.0, 5.4, 5.5, 5.4 | 5.1, 4.5, 3.8, 5.4 | 4.2, 3.7, 5.4, 3.6 | 5.7, 3.2, 5.1, 4.5 | 2.7, 4.7, 5.4, 6.5 |
| 4.5, 3.6, 6.0, 5.0 | 4.9, 4.7, 5.4, 4.5 | 4.6, 2.7, 4.7, 5.1 | 3.8, 5.0, 5.4, 3.9 | 4.9, 6.2, 5.0, 3.6 |
| 4.3, 4.8, 7.0, 3.8 | 4.0, 3.2, 3.9, 5.5 | 4.1, 4.2, 4.8, 3.5 | 4.2, 3.2, 2.5, 3.7 | 3.4, 3.7, 2.9, 5.1 |

Supposing that the two percentiles, $x_R$ and $y_R$, are independent random variables, with the same $\beta$ as it is usually found. The anticipated value for $\beta$ is 5, as before, and the anticipated value of the ratio $u = x_R/y_R$ is assumed to be 0.76 from the ratio of the previous anticipated mean values $\bar{x}_R/\bar{y}_R = 2.9/3.8 \cong 0.76$.

The ratios $\hat{x}_R/\hat{y}_R$ of the percentiles estimated for the two processes give all the twenty-five points of the ratio chart shown in Figure 1. Using the inverse of the transformation (19) and given the false alarm risk, $\alpha = 0.27\,\%$, we obtain the control limits $\text{LCL} = u_{R,\alpha/2}$ and $\text{UCL} = u_{R,1-\alpha/2}$ as simple transformations of the percentiles, $v_{\alpha/2}$ and $v_{1-\alpha/2}$ respectively, of the Inverted Beta. The *prior* $\text{pdf}\{u|\underline{x}_k,\underline{y}_k,\beta\}$ is obtained from the equation (18) by setting $k = 0$.

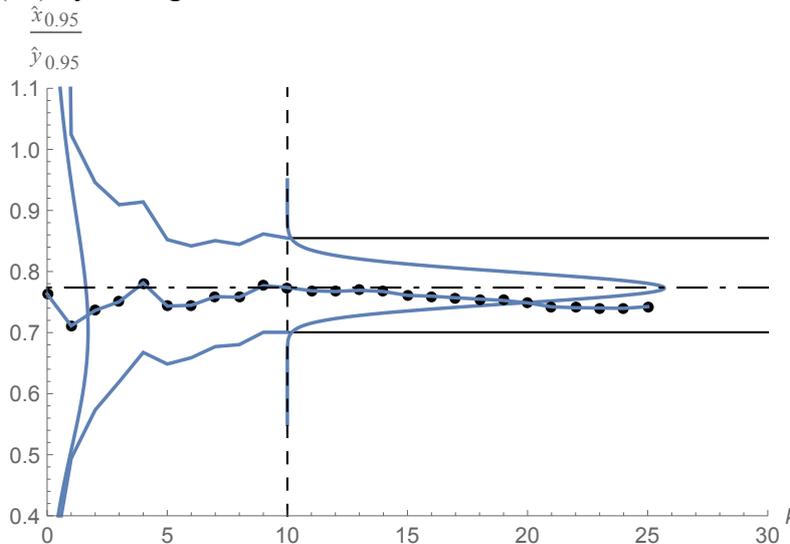

Figure 1. Control Chart for the percentile ratio of the $x$ and $y$ processes; diagram of the $\text{pdf}\{u|\underline{x}_k,\underline{y}_k,\beta\}$ at beginning and end of the Phase I.

Since the ratio chart of Figure 1 does not signal any out-of-control state, we can conclude that the ratio $x_R/y_R$ of the percentiles of the two processes is in statistical control, that is the hypothesis of *homogeneity* of the two productions can be supported.

On the contrary, when the chart signals an out-of-control state, a warning is raised and investigations for assignable causes can start for both the single processes. Obviously, the eventual out-of-control states of both the two single processes can cause an out-of-control state of their ratio only if they result in *non-proportional* effects. Besides, the technical investigation can also include the control charts of the parameters $\beta_x$ and $\beta_y$, to verify the existence of shifts and/or differences in the Weibull shape parameters.

By multiplying by 1.15 the original value of the last fifteen data of Table 2, we simulated an out-of-control state. That implies a 15% positive shift of the $y$ mean from about 4.5 to about 5.2, after the Phase I. The chart designed from the original first ten samples (see, Table 2 and Figure 1) is now used for the new simulated data by setting up the new control chart reported in Figure 2. Obviously, the first ten points of both charts coincide. As previously, the vertical dashed line corresponds to the last in-control sample before the out-of-control state occurs. The chart provides a prompt response at the 12[th] sample after the simulated shift. Moreover, we verified that the simulated shift of $y$ did not affect the in-control state of the Weibull shape parameter $\beta_y$.

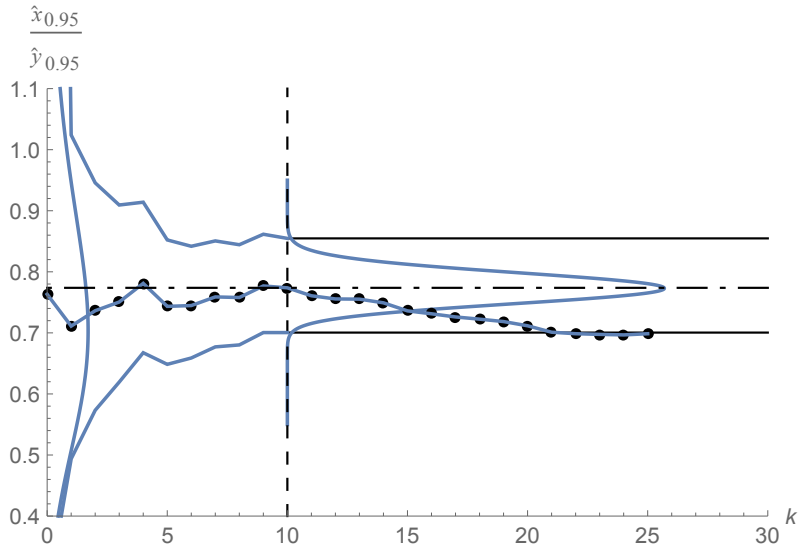

Figure 2. Control Chart for the percentile $x_R/y_R$ after a simulated shift of the $y$ mean from about 4.5 to about 5.2 at the end of the Phase I.

## 4. Example: control of the concrete compressive strength

In the concrete industry, new technical norms have been promulgated to guarantee that the actual quality of the produced concrete matches the prescriptions. Thus, effective quality control techniques, specifically addressed to detect causes, which can affect concrete quality, have become mandatory. This control strategy can also help the manufacturer to reduce costs by checking as soon as possible the compliance with the prescriptions and/or the homogeneity of the product coming from more production lines. Sometimes, the latter characteristic is the main and very critical one, since often the product coming from more production lines must be combined to build the same structure. The parameter of interest is the 0.05 percentile of the distribution of the compressive strength (CS), measured in GPa from the destructive tests of at least two standard sampled specimens. Since the concrete is a quasi-brittle material, the distribution of the CS is generally modelled as a Weibull distribution, with low $\beta$ parameter, say $\beta = 2$. In fact $\beta$ is related to the dispersion of flaws in the material (e.g. see Padgett et al. 1995) and can be considered constant (Nelson 1979). Specifically, due to its intrinsic inhomogeneity, the concrete shows a high variability in strength and so a low $\beta$ value.

Table 3. Compressive strength CS (GPa×10) of standard specimens of produced concrete.

| a) First production line | | | | b) Second production line | | | |
|---|---|---|---|---|---|---|---|
| 3.5, 3.4 | 3.3, 3.3 | 2.9, 2.9 | 4.0, 3.9 | 2.9, 2.9 | 2.9, 3.1 | 2.9, 3.0 | 2.9, 3.0 |
| 3.1, 3.1 | 3.2, 3.1 | 2.9, 3.3 | 2.9, 2.9 | 2.9, 3.0 | 3.1, 2.9 | 3.8, 4.0 | 3.4, 3.6 |
| 3.2, 3.3 | 2.9, 3.0 | 3.3, 3.2 | 3.0, 2.7 | 3.4, 3.5 | 3.3, 3.4 | 2.9, 2.9 | 3.2, 3.1 |
| 2.8, 3.0 | 2.9, 2.9 | 3.0, 2.9 | 2.9, 2.9 | 3.8, 3.4 | 3.2, 3.8 | 2.9, 3.1 | 3.2, 3.4 |
| 2.9, 3.0 | 3.6, 3.6 | 3.1, 3.0 | 3.1, 3.0 | 3.3, 3.0 | 2.8, 2.9 | 3.5, 3.6 | 3.0, 3.0 |
| 4.0, 3.9 | 3.7, 3.6 | | | 3.6, 3.5 | 3.6, 3.6 | | |
| 3.3, 2.9 | 3.4, 3.9 | 2.9, 2.9 | 3.9, 3.3 | 4.0, 2.9 | 3.2, 3.0 | 3.2, 2.9 | 2.9, 2.8 |
| 3.3, 3.0 | 2.9, 3.0 | 2.7, 2.7 | 3.1, 3.2 | 3.4, 2.9 | 2.9, 3.2 | 2.9, 3.3 | 3.6, 2.9 |
| 2.9, 3.0 | 2.9, 2.9 | 3.6, 3.3 | 3.5, 3.0 | 3.8, 3.4 | 2.9, 3.2 | 2.9, 3.2 | 3.5, 3.1 |
| 2.9, 3.5 | 2.9, 3.0 | 3.5, 2.9 | 3.4, 3.0 | 2.9, 2.9 | 3.1, 2.9 | 3.4, 3.4 | 2.9, 3.5 |
| 3.2, 2.9 | 2.9, 3.0 | 2.9, 3.3 | 4.0, 3.3 | 3.2, 3.0 | 2.9, 2.9 | 2.9, 3.8 | 2.9, 3.1 |
| 3.1, 3.1 | 2.8, 3.6 | | | 3.1, 3.2 | 3.0, 3.2 | | |

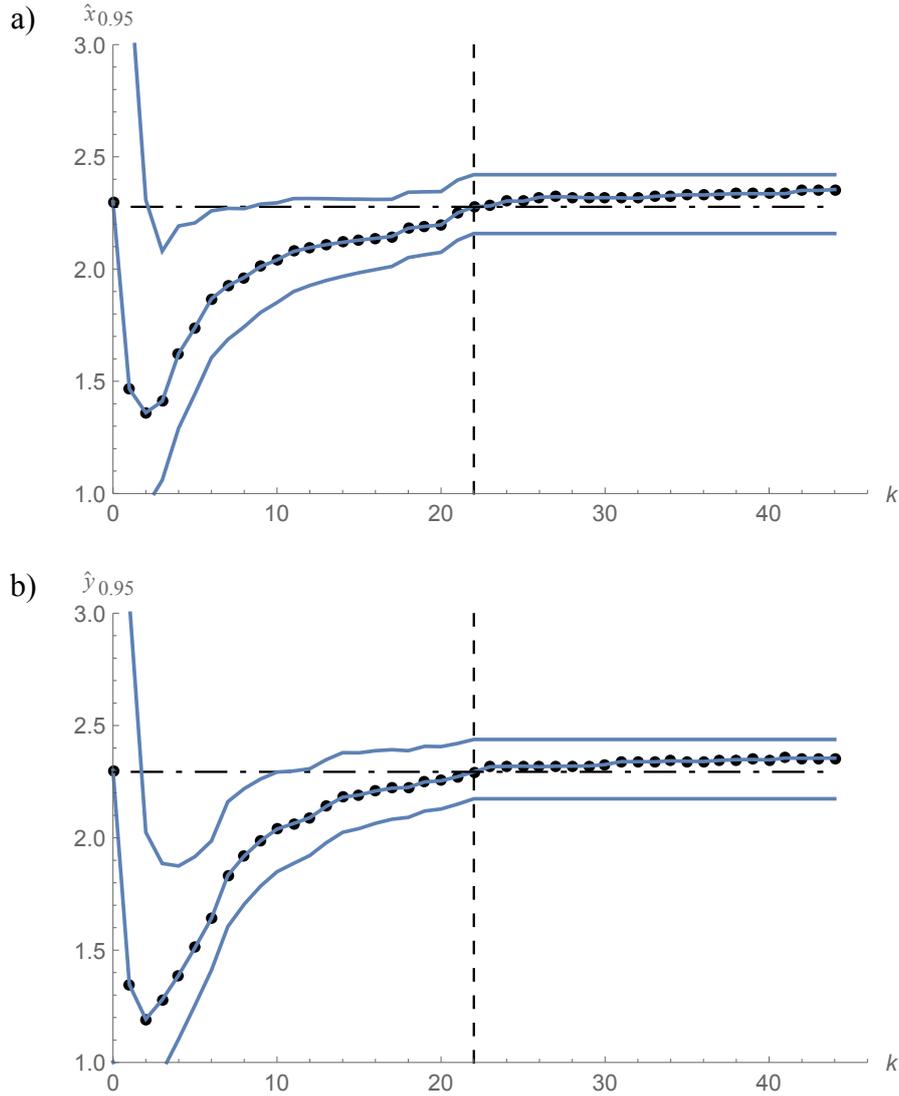

Figure 3. Control Charts for the Weibull percentile of the first a) and second b) process.

From two production lines, a sample of $n = 2$ standard specimens is selected periodically, and the CS of each specimen is measured in GPa (Giga-Pascals) and reported in Table 3 divided by 10. For these specimens the value $\bar{x}_{0.95} = \bar{y}_{0.95} = 2.3$ (GPa×10) is expected to comply with the specifications and so anticipated in (5); the prior interval $2.4 \times (1 \mp 0.5)$ (i.e.: 1.2, 3.6) is anticipated for $\beta$. Figure 3a) and Figure 3b) show the Bayesian control charts obtained using the forty four samples (of size $n = 2$) reported in Table 3a) and in Table 3b) respectively. The first twenty two points of both charts, recursively estimated by (12), correspond to the *tuning* period needed to comply with the specifications and then to achieve the in-control state of the two processes. Consequently, in this case, the Phase I is unusually reduced to very few historical *in-control* data, say about 18 instead of the typical 40÷50 data.

The chart shows to be able to face also this unusual case by identifying the in-control process parameters as accurately as possible, even if very few *training* data are really available.

Given a Shewhart-type false alarm risk $\alpha = 0.27\%$, the UCL and the LCL are calculated by (17). The charts designed from the first twenty-two samples are then used for the next twenty-two samples. The vertical dashed line corresponds to the last sample of the first set.

Supposing that the two percentiles, $x_R$ and $y_R$, are independent random variables, with the same $\beta$ as it is desired, since this is a sign of homogeneity of the product from the two lines. The anticipated value for $\beta$ is 2.4, as before, and the anticipated value of the ratio $\bar{u} = \bar{x}_R / \bar{y}_R$ is assumed to be 1 from the ratio of the previous anticipated mean values $\bar{x}_R / \bar{y}_R = 1$. The ratios $\hat{x}_R / \hat{y}_R$ of the percentiles estimated for the previous two charts give all the forty-four points of the ratio chart shown in Figure 4. Using the inverse of the transformation (19) and given always the same false alarm risk, $\alpha = 0.27\%$, we obtain the control limits $\text{LCL} = u_{R,\alpha/2}$ and $\text{UCL} = u_{R,1-\alpha/2}$ as simple transformations of the percentiles, $v_{\alpha/2}$ and $v_{1-\alpha/2}$ respectively, of the Inverted Beta. For $k = 0$, we obtain the *prior* upper and lower limits. Analogously, the prior $\text{pdf}\{u | \underline{x}_k, \underline{y}_k, \beta\}$ is obtained from the equation (18) by setting $k = 0$.

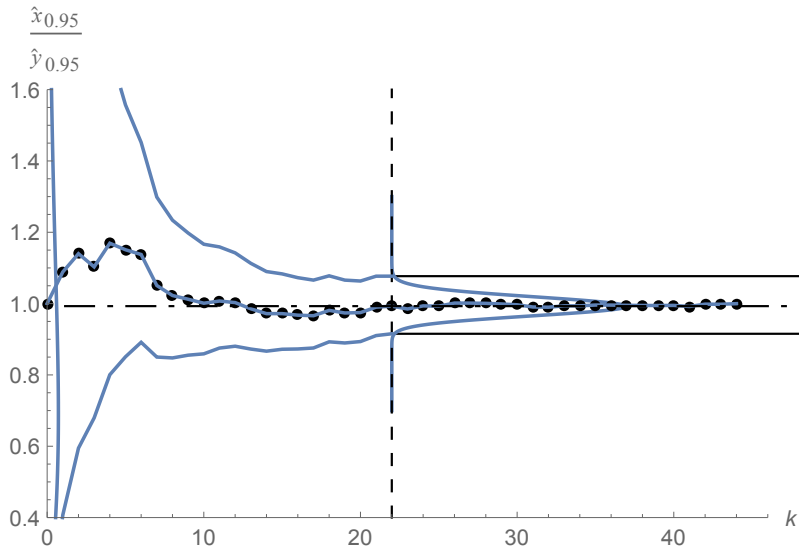

Figure 4. Control Chart for the percentile ratio of the previous two processes (Figure 3); diagram of the $\text{pdf}\{u | \underline{x}_k, \underline{y}_k, \beta\}$ at beginning and end of the *tuning* period.

At the end of the *tuning* period, both charts of Figure 3 show conformity to the prescribed percentile value. No charts signal an out-of-control state. Obviously, this does not imply necessarily the homogeneity of the concrete produced from the two sister lines. So, the control of the percentiles ratio has been needed too. Moreover, this control is useful to point out if the initial marked discrepancy between prior information and data of both processes imply or not their significant dis-homogeneity. Since the ratio chart of Figure 4 does not signal any out-of-control state, the hypothesis of *homogeneity* of the two productions can be supported.

## 5. Conclusion

The primary aim of this note is to highlight that the problem of monitoring the ratio of two processes, which have to comply with some given specifications, can be actually faced via a Bayesian approach in real industrial contexts. For example, if $X$ represents the diameter of a shaft (from a first production line) and $Y$ represents the diameter of a bearing (from a second production line), which is to be mounted upon the shaft, then the basic specification is $X/Y < 1$. To face this problem, it is proposed a Bayesian chart that exploits such specifications as part of the knowledge needed to define the prior pdf of the ratio of the Weibull percentiles of two processes.

The proposed point estimator of the ratio works as a modifier that refines (in mean) the *specified* prior or the *anticipated* prior limits of the ratio of percentiles, approaching rapidly their *posterior actual-current* values. Monitoring of Weibull percentiles is run with no data transformation needed. Even in the case of very small sample sizes, the control chart enables decision-making. The chart can still start from a limited number of samples. In addition, it has the potential to be used for individual observations too. Lastly, the Bayesian framework also provides some additional perspectives such as the computation of the posterior predictive density function that is useful to assess the process capability exactly.

In two applicative examples, the chart complies with the expectations, providing the needed information. However, a complete knowledge of the performance of the chart needs further investigation, even by using a large Monte Carlo study that will find room in a dedicated next paper.